\documentclass[journal]{IEEEtran}
\usepackage{amsfonts,epsfig,cite,array,graphicx,amsmath,amssymb,epstopdf}
\usepackage{caption, longtable, makecell}
\usepackage{multirow}
\usepackage{amsmath}
\usepackage{booktabs}
\usepackage{caption}
\usepackage{capt-of}
\usepackage{balance}
\usepackage{subfigure}
\usepackage{fixltx2e}
\usepackage{setspace}
\usepackage{color, colortbl}
\usepackage{adjustbox}
\usepackage{hyperref}
\usepackage{mathtools}
\usepackage{svg}
\usepackage{subcaption}
\usepackage{graphicx}

\usepackage{algorithm}
\usepackage{algpseudocode} 

\begin{document}

\title{

Exploring Brain Network Organization in Alzheimer’s Disease
and Frontotemporal Dementia: A Crossplot Transition Entropy Approach

}


\author{{Shivani Ranjan, Lalan Kumar}
\thanks{Shivani Ranjan is with 
the Department of Electrical Engineering, IIT Delhi, New Delhi, India (e-mail: eez208482@iitd.ac.in).}
\thanks{ Lalan Kumar is with the Department of Electrical Engineering, Bharti School of Telecommunication, and Yardi School of Artificial Intelligence, Indian Institute of Technology Delhi, New Delhi 110016, India (e-mail: lkumar@ee.iitd.ac.in).}}

\maketitle

\begin{abstract}

Dementia poses a growing challenge in our aging society. Frontotemporal dementia (FTD) and Alzheimer’s disease (AD) are the leading causes of early-onset dementia. FTD and AD display unique traits in their onset, progression, and treatment responses. In particular, FTD often faces a prolonged diagnostic process and is commonly misdiagnosed with AD due to overlapping symptoms. This study utilizes a complex network model of brain electrical activity using resting-state EEG recordings to address the misdiagnosis. It compares the network organization between AD and FTD, highlighting connectivity differences and examining the significance of EEG signals across frequency bands in distinguishing AD and FTD. The publicly available EEG dataset of 36 AD and 23 FTD patients is utilized for analyses. Cross-plot transition entropy (CPTE) is employed to measure synchronization between EEG signals and construct connection matrices. CPTE offers advantages in parameter setting, computational efficiency, and robustness. The analysis reveals significantly different clustering coefficients (CC), subgraph centrality (SC), and eigenvector centrality (EC) between the two groups. FTD shows higher connectivity, particularly in delta, theta, and gamma bands, owing to lower neurodegeneration. The CPTE-based network parameters effectively classify the two groups with an accuracy of 87.58\%, with the gamma band demonstrating the highest accuracy of 92.87\%. Consequently, CPTE-based, complex network analysis of EEG data from AD and FTD patients reveals significant differences in brain network organization. The approach shows potential for identifying unique characteristics and providing insights into the underlying pathophysiological processes of the various forms of dementia, thereby assisting in accurate diagnosis and treatment.

\end{abstract}


\section{Introduction}
\label{sec:introduction}

\begin{table*}[h!]
\caption{ Table represents the demographic information of the data}

\centering

    \scalebox{0.8}{
\begin{tabular}{|c|c|c|c|c|c|c|c|}
\hline
\textbf{Group} &
  \textbf{N} &
  \textbf{MMSE\_{mean}} &
  \textbf{MMSE\_{std.dev}} &
  \textbf{Age\_{mean}} &
  \textbf{Age\_std.dev} &
  \textbf{Disease\_duration\_median} &
  \textbf{Disease\_duration\_IQR} \\ \hline
AD  & 36 & 17.75 & 4.5  & 66.4 & 7.9 & 25 & 24 \\ \hline
FTD & 23 & 22.17 & 8.22 & 63.6 & 8.2 & 25 & 24 \\ \hline
HC  & 29 & 30    & 0    & 67.9 & 5.4 & 25 & 24 \\ \hline
\end{tabular}}
\end{table*}

The incidence of dementia is on the rise, driven by the global aging population \cite{prince2016world}. Alzheimer’s disease (AD) stands out as the most prevalent form of dementia, comprising an estimated 60$\%$ of all cases \cite{prince2016world}. The pervasive prevalence of AD coupled with the absence of approved disease treatments, significantly impact patients' daily functioning and quality of life \cite{feigin2019global}. The cognitive impairments associated with AD remain inadequately understood \cite{cosentino2005metacognitive}. In particular, frontotemporal dementia (FTD) is frequently misdiagnosed as AD \cite{rosso2003frontotemporal}. This misdiagnosis arises from shared clinical symptoms, including behavioral changes, executive dysfunction, language difficulties, and motor impairments \cite{ratnavalli2002prevalence,leroy9characteristics}. FTD, a prevalent form of dementia, carries inherent safety risks for affected individuals \cite{bang2015frontotemporal}. Decision-making activities, including medication management and meal preparation, carry potential risks leading to severe consequences \cite{delozier2016systematic}. Achieving an accurate diagnosis of AD and FTD, crucial for improving patient treatment outcomes, necessitates advancements in current diagnostic tools.

The shift to EEG investigations is motivated by the urgent need for timely and accurate diagnoses \cite{oltu2021novel} for patients with FTD. They experience a more rapid cognitive decline and shorter survival rates compared to those with AD \cite{rascovsky2011sensitivity}. Previous research has explored cognitive aspects in FTD and AD patients \cite{massimo2012self} using structural imaging. It highlights differences in self-appraisal and grey matter density. The meta-cognitive assessments \cite{cosentino2005metacognitive} have pinpointed specific cognitive deficits in AD and FTD. Notably, FTD patients exhibit greater monitoring disorders than AD patients.
Although metacognition measures show promise in distinguishing between AD and FTD, they come with limitations such as cost, time-consuming procedures, and subjectivity \cite{cassani2017towards}. It is crucial to note the scarcity of promising tools that can effectively diagnose and differentiate between AD and FTD based on neural evidence.

EEG signals in neurodegenerative patients exhibit distinctive features compared to those in elderly and healthy controls. These features include the slowing effect and synchrony between pairs of EEG signals \cite{adeli2005alzheimer}. These effects are believed to stem from disrupted segregation and integration in brain networks due to functional disconnection caused by neuronal death \cite{adeli2005alzheimer}. Morabito et al. \cite{morabito2015longitudinal} reported a decline in functional connectivity in longitudinally evaluated AD patients. Tahaei et al. \cite{tahaei2012synchronizability} found decreased synchronizability across delta, alpha, beta, and gamma EEG frequency bands. De Haan et al. \cite{de2009functional} observed a degeneration of large-scale functional brain network organization in AD patients. However, functional connectivity and network studies in behavioral variant frontotemporal dementia (bvFTD) are limited. Prior research indicates that EEG in bvFTD remains normal or only mildly disturbed until late in the disease progression \cite{neary2005frontotemporal, neary1998frontotemporal, stam2012dementia}. Pijnenburg et al. \cite{pijnenburg2008investigation} found higher functional connectivity in bvFTD compared to AD in the upper alpha band (10–13 Hz). Yu et al. \cite{yu2016different} observed higher synchronization in bvFTD in the delta and alpha bands compared to AD, while in the theta band, AD exhibited higher phase lag index (PLI) than bvFTD. 

The cross-plot transition entropy (CPTE) was recently introduced to measure the synchronization of bivariate time series. This method offers advantages like easy parameter setting, efficiency, robustness, good consistency, and suitability for short time series \cite{chen2023capturing}. In comparison with existing methods such as permutation mutual information (PMI), cross-permutation entropy (CPE), phase lag index (PLI), and joint-order pattern recurrence plot (JORP), the state evaluation of coupling degree of CPTE is nearly independent of parameter settings and exhibits the highest computational efficiency \cite{chen2023capturing}. An analysis based on CPTE was proposed for evaluating EEG data from auditory-evoked potentials. The synchronization between electrode pairs varies significantly during auditory stimulation in CPTE compared to other synchronization measures \cite{chen2023capturing}. Inspired by the low computational cost and success of the CPTE approach in translating neural signal synchronization into various types of auditory stimulations, this study explores the feasibility of using surface EEG to differentiate neurodegenerative diseases.

The neurodegenerative disorder FTD, often has a longer diagnostic journey and is frequently misdiagnosed with AD due to their overlapping clinical symptoms. However, FTD and AD exhibit unique traits in their onset, progression, and response to treatment, which can be captured in neural signals recorded from the scalp. Despite this, there has been limited attention given to differentiating and understanding the brain network organization of neurodegenerative diseases with similar symptoms using short-term information from surface EEG. This is the main focus of this study.
This paper integrates ordinal pattern transition networks \cite{ruan2019ordinal, yang2020multivariate, kulp2016using} into the cross-plot \cite{yang2021using,yang2020multivariate} to compute CPTE. Two cohorts of 36 AD patients and 23 FTD subjects are considered. This study marks the first comparison of resting-state EEG signals between AD and FTD, using CPTE to identify cross-sectional differences between the two groups. In particular, the key contributions of this article are outlined below.

\begin{itemize}
 
    \item Surface EEG-based AD and FTD differentiation using CPTE synchronization measure.
    \item Development of a CPTE-based complex network model utilizing bivariate short-time series.
    \item Analysis of brain network organization for AD and FTD using complex network parameters and connectivity density variation.
    \item EEG frequency band analysis for AD and FTD groups.
\end{itemize}

The paper is structured as follows: Section \ref{material} presents details about EEG data and preprocessing. Section \ref{methodology} outlines the CPTE-based complex network approach proposed to investigate the large-scale brain network organization in AD and FTD. Section (\ref{result}) presents the results, followed by a discussion of the findings in Section \ref{discussion}. Finally, Section \ref{conclusion} draws conclusions based on the study.



\begin{figure*}[!h]
      \centering
      \adjustbox{trim= 0 0cm 0 0}{%
      \includegraphics[width = 1\linewidth]{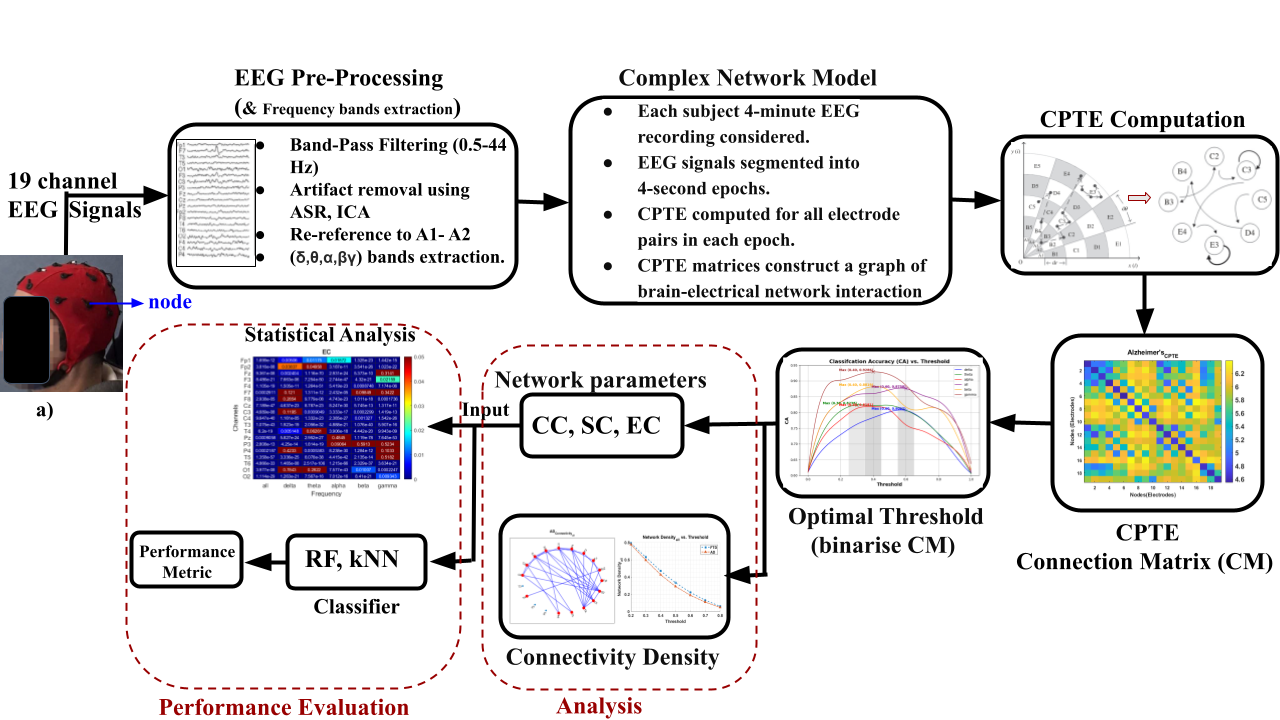}}
      \caption{The block diagram illustrates the methodical process employed to analyze disparities in large-scale brain network organization between neurodegenerative disease groups (AD, FTD) using a CPTE-based complex network model.}
      \label{block_Diagram}
   \end{figure*}

\section{Materials: Data and PreProcessing}
\label{material}

Publicly available resting state EEG data from 88 participants \cite{ds004504:1.0.6} is utilized in this study. The data was recorded with a Nihon Kohden EEG 2100 clinical device. The 19 scalp electrodes were placed according to the 10-20 international system, with two mastoid reference electrodes (A1 and A2) to monitor impedance below 5 $k\Omega$. The signal was sampled at 500 Hz with a resolution of 10 $uV/mm$. 
Rigorous preprocessing in MATLAB and EEGLAB \cite{math2020matlab, delorme2004eeglab} included band-pass filtering (0.5-44 Hz) with a Butterworth filter and re-referencing to A1-A2 \cite{ds004504:1.0.6}. The Artifact Subspace Reconstruction Routine (ASR) \cite{chang2019evaluation} was employed to eliminate non-cerebral artifacts, surpassing conventional thresholds by identifying and removing data periods that significantly improved quality. The EEG data was decomposed into independent sources using independent component analysis (ICA) with "ICLabel" tool in EEGLAB. Artifact-related components (e.g., "eye artifacts" or "jaw artifacts") were automatically identified and excluded thereafter. The process ensured the retention of neural components, enhancing the signal-to-noise ratio. Subsequently, for brainwave analysis across different frequency bands (delta: 0.5–4 Hz, theta: 4–8 Hz, alpha: 8–13 Hz, beta: 13–30 Hz, and gamma: 31–44 Hz), the EEG signals were band-pass filtered using Butterworth filters for each specific frequency range.

\section{Methodology: CPTE based Complex Network Model}
\label{methodology}

In this paper, the potential of EEG-based complex network analysis to capture differences in the brain electrical network organization of AD and FTD subjects is explored as its first objective. For this purpose, a graph representation for the interaction between the electrical activity of different cortex areas is adopted. The nodes of the graphical network are represented by the electrodes that cover the cortical areas, as shown in Figure \ref{block_Diagram} (a). The weight of an edge connecting two nodes is determined by estimating the coupling strength between the corresponding EEG signals. This results in the construction of graph representation for the brain electrical network interaction. This graph model is estimated for every subject (EEG recording). The quantitative analysis of the graph model indirectly provides information about the efficiency of the electrical network organization of the brain in that subject.

A 4-minute recorded EEG signal was taken for further analysis. The data was segmented into thirty segments, each comprising of 4000 samples. The sliding time window method was employed within each segment to facilitate short time series signal analysis. For this purpose, a window of length 2000 samples with a step size of 500 samples was utilized. The coupling strength becomes constant beyond 2000 samples\cite{chen2023capturing}. Thus, for each subject, 150 epochs were analyzed. This leads to 5400 epochs for 36 AD patients and 3450 epochs for 23 FTD patients. For every epoch "$e$," the cross-plot transition entropy (CPTE) is computed next. The CPTE parameter will be utilized for differentiating the AD and FTD patients by generating a graph model.
\subsection{Cross-plot Transition Entropy (CPTE)}
\label{CPTE_Analysis}
\begin{figure*}[!h]
      \centering
      \adjustbox{trim= 0 0cm 0 0}{%
      \includegraphics[width = 1\linewidth]{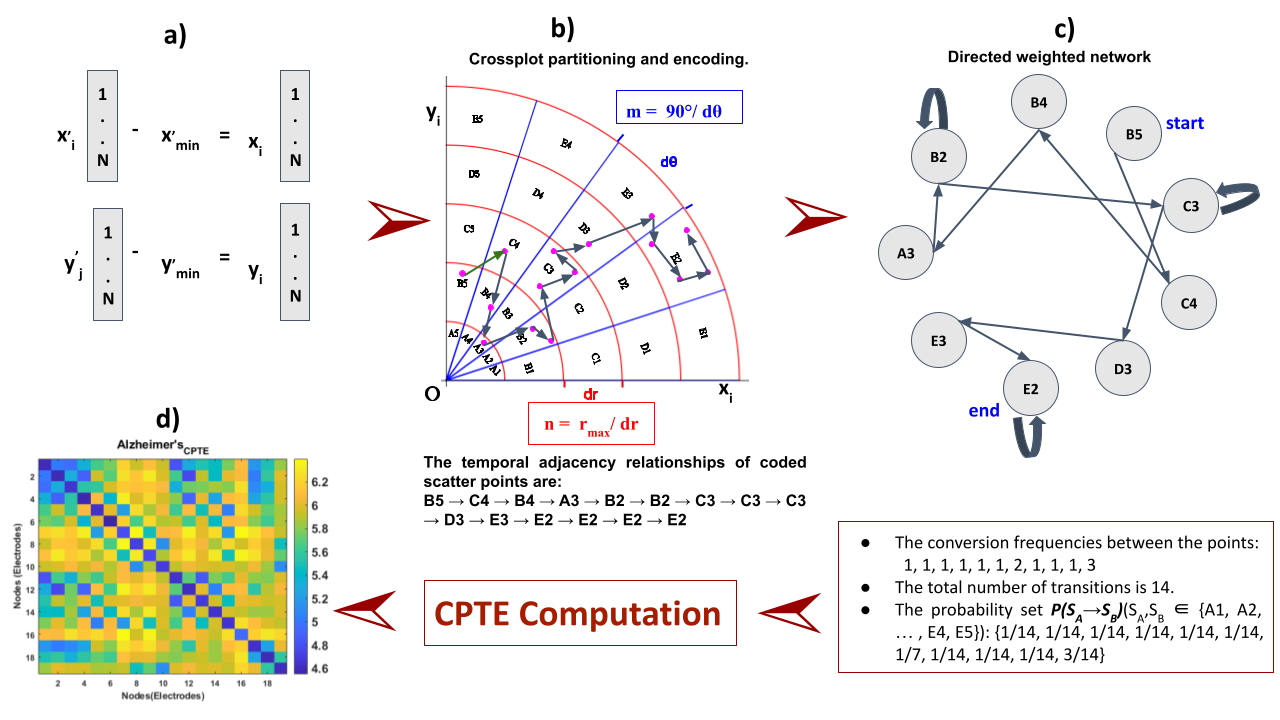}}
      \caption{The diagram illustrates the computation of CPTE, where x'$_i$ and y'$_i$ represent two time series, in this case, an epoch of a possible electrode pair. In this analysis, the radial ruler ($dr$) and angular ruler ($d\theta$) are set to 10 and $10^\circ$, respectively.}
      \label{CPTE_BD1}
   \end{figure*}

The CPTE introduced in \cite{chen2023capturing}, draws inspiration from ordinal pattern transition properties in time series analysis. It quantifies the regularity of these transitions using Shannon entropy, serving as a measure of synchronization between two time series. The computation of CPTE is illustrated in Figure \ref{CPTE_BD1}. A brief discussion is presented herein for completeness. 

From the given two synchronous time series \( x'_i \) and \( y'_i \), a pair of new time series \( x_i \) and \( y_i \) is obtained by subtracting them with their respective minimum values to construct a cross plot with scatter points located in the first quadrant of Cartesian coordinates as shown in Figure \ref{CPTE_BD1}. Here $x_i$ and $y_i$ represent any two EEG sensor time series data with $i=1,2..N$. The cross plot is now partitioned and coded using a method that integrates phase and amplitude information from scatter points with an angular ruler ($d\theta$ = 10$^\circ$) and a radial ruler ($dr$ = 10) parameters. 
Each pair of $(x_i,y_i)$ occupies one of the ring subinterval of sectors forming nodes. Subsequently, a directed weighted network is constructed according to the temporal adjacency relationship
between points, with the number of transitions as the weight of the network. 
A discrete probability set $  P(\text{S}_\text{A} \rightarrow \text{S}_\text{B})$ of
transition behaviors are determined based on the constructed directed weighted network, where \( P(\text{S}_\text{A} \rightarrow \text{S}_\text{B}) \) denotes the probability of transition from \(\text{S}_\text{A}\) to \(\text{S}_\text{B}\), and \(\text{S}_\text{A}\),\(\text{S}_\text{B}\) $\in$ \{ A1, A2, … , E4, E5 \}.
 The CPTE is finally formulated as
\begin{equation}
    CPTE = - \sum P(\text{S}_\text{A} \rightarrow \text{S}_\text{B}) \log_2 P(\text{S}_\text{A} \rightarrow \text{S}_\text{B})  
\end{equation}
It is to note that CPTE value is inversely proportional to the coupling strength between two time series.

For every epoch "$e$," the CPTE$^e$(n$_i$, n$_j$) between every possible pair of electrodes n$_i$ and n$_j$ was calculated. Here, CPTE$^e$(n$_i$, n$_j$) is the (n$_i$, n$_j$)$_{th}$ element of the synchronization matrix CPTE$^e$ of epoch $e$. For $19$ channel EEG data, $171$ possible pairs of channels were considered. Since CPTE$^e$(n$_i$, n$_j$) = CPTE$^e$(n$_j$, n$_i$), the CPTE$^e$ matrix is symmetrical, and the graph model is, therefore, undirected. Once the analysis of the entire recording of a subject is completed, each CPTE$^e$ symmetric matrix is normalized by Min-Max normalization so that the elements of the matrices CPTE$^e$ fall in the range 0–1. The matrix is utilized next for various network parameter computations.

\subsection{Complex Network Measure}
\label{Network_Analysis}

The primary goal is to assess the effectiveness of the complex network organization as it reflects the efficiency of the underlying cortical-electrical network organization. Network efficiency is evaluated for the entire EEG dataset and every epoch $e$. Various network measures such as clustering coefficient (CC), subgraph centrality (SC), and eigenvector centrality (EC), are now computed for each CPTE$^e$ matrix.

The clustering coefficient (CC) measures segregation, indicating the likelihood that two neighbors of node $k$ become neighbors of each other. The average CC is defined as:
 \begin{align}
\label{eq1}
\textrm{CC} &=\frac{1}{n} \sum_{k=1}^{n} CC_k = \frac{1}{n} \sum_{k\in V} \frac{2t_k}{p_k(p_k - 1)}
\end{align}
where CC$_k$ represents the clustering coefficient of node $k$, $n$ indicates the total number of nodes in the network, $V$ reflects the set of all nodes in the network, $t_k$ is the number of triangles that include node $k$, and $\frac{p_k(p_k - 1)}{2}$ calculates the maximum number of possible connections of that node. Here, $p_k$ is the degree of node $k$ and the degree of a node in a weighted network is the sum of the weights of the edges connected to that node.

The subgraph centrality (SC) is a measure of network centrality that assesses a node's involvement in all potential subgraphs of a network. It is defined as:
\begin{align}
\label{eq3}
\textrm{C}_s(l) &= \sum_{l=1}^{\infty} \frac{1}{l!} \times \textrm{Tr}(A^l)
\end{align}
where $A$ represents the CPTE$^e$ matrix, $l$ is the length of closed walks, $Tr(A^l)$ is the trace of $A^{l}$, and the summation is taken over all possible walk lengths.

The eigenvector centrality (EC) measures the importance of a node in a network, considering both the number of connections a node has and the importance of those connections. It is defined as:
 \begin{align}
\label{eq2}
Ax &= {\lambda} x
\end{align}
where $A$ represents the CPTE$^e$ matrix, $x$ is the eigenvector, and $\lambda$ is the associated eigenvalue. The centrality scores are derived by normalizing the final eigenvector. 
Nodes with higher EC scores are deemed more central, signifying their significance in the structure and connectivity of the network.

     \begin{figure}[h!]
    \centering
    \includegraphics[width = 0.48\textwidth]{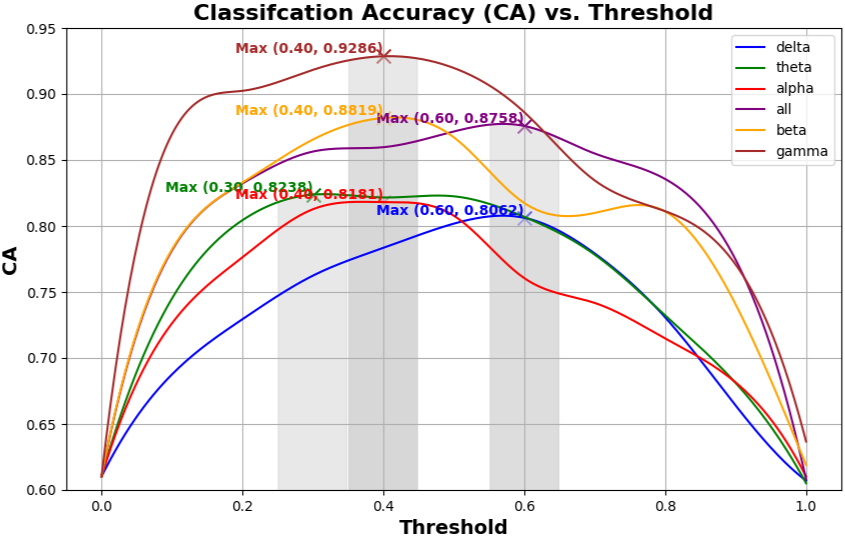}
    \caption{The figure depicts the exploration of optimal threshold values by varying the threshold $th$ from 0 to 1 in steps of 0.1. The threshold with the highest classification accuracy of Random Forest (RF) in distinguishing between AD and FTD is selected for analyzing the differences in brain network organization.}
    \label{CA_threshold}
\end{figure}

The network efficiency parameters described are utilized as features for CPTE\(^e\) connection matrix nodes (electrodes). This results in $19 \times np$ features for each epoch $e$, where $np$ equals 1 for individual network parameters (CC, SC, or EC) and 3 for all network parameters combined. To evaluate the performance of these features in classifying neurodegenerative diseases (AD and FTD), classification accuracy is assessed using a 10-fold cross-validation technique. Initially, the optimal threshold value is explored by varying the threshold $th$ (ranging from 0 to 1 with a step of 0.1) and observing the classification accuracy of Random Forest (RF) in discriminating between AD and FTD. Subsequently, the performance of the alternative classifier k-Nearest Neighbors (kNN) is assessed at the optimal threshold value using its default parameters. The statistical significance of each feature in discriminating between the AD and FTD groups is evaluated using a t-test. Additionally, insights into AD and FTD differentiation across all frequency bands of the EEG signal (delta: 0.5–4 Hz, theta: 4–8 Hz, alpha: 8–13 Hz, beta: 13–30 Hz, and gamma: 31–44 Hz) are explored using the same methodology.

The network efficiency parameters are utilized as features for nodes (electrodes), resulting in a total of $19 \times 3$ features for each epoch $e$. Initially, we explore the optimal threshold value by varying the threshold and observing the classification accuracy of Random Forest (RF) in distinguishing between AD and FTD. Subsequently, we assess the performance of an alternative classifier, k-Nearest Neighbors (kNN), using its default parameters. Additionally, we conduct t-tests to determine the statistical significance of each feature in discriminating between the AD and FTD groups.


\begin{table*}[h!]
\caption{The table illustrates the disparity in median values of network efficiency parameters between the AD and FTD groups, accompanied by the corresponding statistical significance denoted by p-values from the t-test.}
\label{feature_descrip}
\centering

    \scalebox{0.8}{

\begin{tabular}{|c|cc|cc|cc|cc|cc|cc|}
\hline
\textbf{bands}  & \multicolumn{2}{c|}{\textbf{all (0.5-44Hz)}}                                      & \multicolumn{2}{c|}{\textbf{delta (0.5-4Hz)}}                                     & \multicolumn{2}{c|}{\textbf{theta (4-8Hz)}}                                       & \multicolumn{2}{c|}{\textbf{alpha (8-13Hz)}}                                      & \multicolumn{2}{c|}{\textbf{beta (13-30Hz)}}                                      & \multicolumn{2}{c|}{\textbf{gamma (31-44Hz)}}                                     \\ \hline
                & \multicolumn{1}{c|}{\textbf{Median}} & \textbf{p-value}                           & \multicolumn{1}{c|}{\textbf{Median}} & \textbf{p-value}                           & \multicolumn{1}{c|}{\textbf{Median}} & \textbf{p-value}                           & \multicolumn{1}{c|}{\textbf{Median}} & \textbf{p-value}                           & \multicolumn{1}{c|}{\textbf{Median}} & \textbf{p-value}                           & \multicolumn{1}{c|}{\textbf{Median}} & \textbf{p-value}                           \\ \hline
\textbf{CC-AD}  & \multicolumn{1}{c|}{0.5813}          & \multirow{2}{*}{\textbf{\textless{}0.001}} & \multicolumn{1}{c|}{0.6258}          & \multirow{2}{*}{\textbf{\textless{}0.001}} & \multicolumn{1}{c|}{0.3305}          & \multirow{2}{*}{\textbf{0}}                & \multicolumn{1}{c|}{0.5689}          & \multirow{2}{*}{0.2252}                    & \multicolumn{1}{c|}{0.5115}          & \multirow{2}{*}{\textbf{\textless{}0.001}} & \multicolumn{1}{c|}{0.4618}          & \multirow{2}{*}{\textbf{\textless{}0.001}} \\ \cline{1-2} \cline{4-4} \cline{6-6} \cline{8-8} \cline{10-10} \cline{12-12}
\textbf{CC-FTD} & \multicolumn{1}{c|}{0.572}           &                                            & \multicolumn{1}{c|}{0.6606}          &                                            & \multicolumn{1}{c|}{0.4351}          &                                            & \multicolumn{1}{c|}{0.5623}          &                                            & \multicolumn{1}{c|}{0.4556}          &                                            & \multicolumn{1}{c|}{0.4522}          &                                            \\ \hline
\textbf{SC-AD}  & \multicolumn{1}{c|}{0.0487}          & \multirow{2}{*}{\textbf{\textless{}0.001}} & \multicolumn{1}{c|}{0.4820}          & \multirow{2}{*}{\textbf{\textless{}0.001}} & \multicolumn{1}{c|}{2.3505}          & \multirow{2}{*}{\textbf{\textless{}0.001}} & \multicolumn{1}{c|}{0.0658}          & \multirow{2}{*}{0.5941}                    & \multicolumn{1}{c|}{0.0314}          & \multirow{2}{*}{\textbf{0.02}}             & \multicolumn{1}{c|}{0.2518}          & \multirow{2}{*}{\textbf{\textless{}0.001}} \\ \cline{1-2} \cline{4-4} \cline{6-6} \cline{8-8} \cline{10-10} \cline{12-12}
\textbf{SC-FTD} & \multicolumn{1}{c|}{0.2290}          &                                            & \multicolumn{1}{c|}{1.8173}          &                                            & \multicolumn{1}{c|}{10}              &                                            & \multicolumn{1}{c|}{0.1222}          &                                            & \multicolumn{1}{c|}{0.0679}          &                                            & \multicolumn{1}{c|}{3.0206}          &                                            \\ \hline
\textbf{EC-AD}  & \multicolumn{1}{c|}{0.6145}          & \multirow{2}{*}{\textbf{\textless{}0.001}} & \multicolumn{1}{c|}{0.7731}          & \multirow{2}{*}{\textbf{0.03}}             & \multicolumn{1}{c|}{0.8225}          & \multirow{2}{*}{\textbf{\textless{}0.001}} & \multicolumn{1}{c|}{0.5781}          & \multirow{2}{*}{\textbf{\textless{}0.001}} & \multicolumn{1}{c|}{0.5978}          & \multirow{2}{*}{\textbf{\textless{}0.001}} & \multicolumn{1}{c|}{0.6939}          & \multirow{2}{*}{\textbf{\textless{}0.001}} \\ \cline{1-2} \cline{4-4} \cline{6-6} \cline{8-8} \cline{10-10} \cline{12-12}
\textbf{EC-FTD} & \multicolumn{1}{c|}{0.6824}          &                                            & \multicolumn{1}{c|}{81.2}           &                                            & \multicolumn{1}{c|}{0.8684}          &                                            & \multicolumn{1}{c|}{0.609}           &                                            & \multicolumn{1}{c|}{0.5943}          &                                            & \multicolumn{1}{c|}{0.723}           &                                            \\ \hline
\end{tabular}

}
\end{table*}

\begin{table*}[ht!]
 \caption{ The table presents the performance metrics of network parameters derived from the CPTE connection matrix, which is binarized at the optimal threshold for each case. These metrics assess the ability to differentiate between AD and FTD.}
 \label{CA}
    \centering
    \scalebox{0.58}{

\begin{tabular}{llllllllllllll}
\hline
\textbf{}                                  & \textbf{}           & \textbf{}                             & \textbf{CC}                           & \textbf{}                             & \textbf{}                             & \textbf{SC}                           & \textbf{}                             & \textbf{}                            & \textbf{EC}                          & \textbf{}                            & \textbf{}         & \textbf{All}         & \textbf{}                             \\ \hline
\textbf{Bands}                             & \textbf{Classifier} & \textbf{Accuracy}                     & \textbf{Sensitivity}                  & \textbf{Specificity}                  & \textbf{Accuracy}                     & \textbf{Sensitivity}                  & \textbf{Specificity}                  & \textbf{Accuracy}                    & \textbf{Sensitivity}                 & \textbf{Specificity}                 & \textbf{Accuracy} & \textbf{Sensitivity} & \textbf{Specificity}                  \\ \hline
                                           & RF                  & 84.23±1.47                         & 68.19±2.37                         & 94.45±1.51                         & 85.85±0.66                         & 73.62±2.53                         & \cellcolor[HTML]{FFFFFF}93.61±1.44 & {\color[HTML]{000000} 85.02±1.41} & {\color[HTML]{000000} 72.06± 2.78} & {\color[HTML]{000000} 93.27±1.01} & 87.58±0.67     & 75.19±1.97        & 95.47±0.88                         \\ \cline{2-14} 
\multirow{-2}{*}{\textbf{all (0.5-44Hz)}}    & Knn                 & \cellcolor[HTML]{FFFFFF}78.81±1.23 & \cellcolor[HTML]{FFFFFF}70.55±2.06 & \cellcolor[HTML]{FFFFFF}84.07±1.99 & \cellcolor[HTML]{FFFFFF}78.95±1.78 & \cellcolor[HTML]{FFFFFF}68.37±2.27 & \cellcolor[HTML]{FFFFFF}85.69±1.50 & {\color[HTML]{000000} 83.06±1.67} & {\color[HTML]{000000} 72.12±2.46} & {\color[HTML]{000000} 90.07±1.52} & 82.21±0.77     & 75.51±2.37        & 86.45±0.88                         \\ \hline \hline
                                           & RF                  & 76.56±1.46                         & 50.92±2.69                         & 92.95±1.39                         & 78.80±1.12                         & 59.14±2.46                         & 91.33±1.00                         & {\color[HTML]{000000} 79.91±1.26} & {\color[HTML]{000000}61.69±1.84} & {\color[HTML]{000000} 91.52±1.46} & 80.62±1.13     & 60.02±1.82        & \cellcolor[HTML]{FFFFFF}93.76±1.21 \\ \cline{2-14} 
\multirow{-2}{*}{\textbf{delta (0.5-4Hz)}}   & Knn                 & 72.90±1.04                        & 62.45±2.07                         & 79.63±2.34                         & 71.67±0.98                         & 56.45±1.37                         & 81.38±1.43                         & {\color[HTML]{000000} 78.62±1.17} & {\color[HTML]{000000} 68.21±1.79} & {\color[HTML]{000000} 85.23±1.85} & 76.89±1.12     & 67.51±2.53        & 82.89±1.47                         \\ \hline \hline
                                           & RF                  & 78.88±1.12                         & 57.90±1.07                         & 92.27±0.97                         & 76.98±1.04                         & 56.49±3.62                         & 90.03±1.07                         & {\color[HTML]{000000} 0.81.21±1.07} & {\color[HTML]{000000} 0.63.45±2.12} & {\color[HTML]{000000} 92.53±0.81} & 82.38±0.75     & 65.20±1.68        & 93.35±0.99                         \\ \cline{2-14} 
\multirow{-2}{*}{\textbf{theta (4-8 Hz)}}  & Knn                 & 79.01±1.80                         & 71.89±1.78                         & 83.57±2.23                         & 74.34±1.61                         & 61.79±2.02                         & 82.37±1.39                         & {\color[HTML]{000000} 81.77±0.75} & {\color[HTML]{000000} 74.01±1.70} & {\color[HTML]{000000} 86.74±1.09} & 83.60±1.23     & 75.96±2.05        & 88.49±2.00                         \\ \hline \hline
                                           & RF                  & 78.41±1.46                         & 57.23±2.31                         & 91.97±1.80                         & 79.42±0.86                          & 61.10±1.37                         & 91.13±1.08                         & {\color[HTML]{000000} 79.13±0.89} & {\color[HTML]{000000} 61.41±1.74} & {\color[HTML]{000000} 90.46±1.69} & 81.81±0.87     & 64.01±0.46        & 93.17±0.05                         \\ \cline{2-14} 
\multirow{-2}{*}{\textbf{alpha (8-13 Hz)}} & Knn                 & 72.67±1.61                         & 63.32±2.57                         & 78.66±1.38                         & 71.94±1.43                         & 60.49±2.89                         & 79.27±1.24                         & {\color[HTML]{000000} 76.89±1.09} & {\color[HTML]{000000} 68.00±2.36} & {\color[HTML]{000000} 82.55±1.66} & 76.01±0.71     & 67.96±1.68        & 81.17±1.10                         \\ \hline \hline
                                           & RF                  & 86.51±0.75                         & 72.00±1.64                         & 95.81±0.74                         & 87.45±1.29                         & 76.10±2.45                         & 94.70±0.76                         & {\color[HTML]{000000} 85.03±1.48} & {\color[HTML]{000000} 69.91±2.56} & {\color[HTML]{000000} 94.72±0.72} & 88.28±1.66     & 75.77±3.41        & 96.31±0.61                         \\ \cline{2-14} 
\multirow{-2}{*}{\textbf{beta (13-30Hz)}}  & Knn                 & 82.32±0.99                         & 79.17±1.55                         & 84.32±0.84                         & 79.28± 1.53                         & 70.17±3.75                         & 85.05±1.12                         & {\color[HTML]{000000} 83.97±1.39} & {\color[HTML]{000000} 76.97±2.12} & {\color[HTML]{000000} 88.46±1.14} & 84.52±1.29     & 82.12±2.08        & 86.08±1.08                         \\ \hline \hline
                                           & RF                  & 90.67±0.77                         & 80.91±2.19                         & 96.93±0.82                         & 91.21±1.36                         & 83.55±2.43                         & 96.15±0.96                         & {\color[HTML]{000000} 92.41±0.76} & {\color[HTML]{000000} 85.04± 1.62} & {\color[HTML]{000000} 97.14±0.76} & 92.87±0.99     & 85.95±2.37        & 97.34±0.58                         \\ \cline{2-14} 
\multirow{-2}{*}{\textbf{gamma (31-44Hz)}} & Knn                 & 87.08±1.17                         & 84.50±1.93                         & 88.77±1.23                         & 86.38±0.90                         & 81.21±2.13                         & 89.75±0.86                         & {\color[HTML]{000000} 92.46±0.76} & {\color[HTML]{000000} 90.44±1.17} & {\color[HTML]{000000} 93.76±0.78} & 89.66±0.92     & 87.75±1.32        & 90.92±1.17                         \\ \hline
\end{tabular}

}
\end{table*}
\subsection{Connectivity density}
\label{connectivity_analysis}
The analysis of connectivity density is performed by examining the averaged matrix CPTE across epochs and cases (AD, FTD). For a given CPTE$^e$, the complementary connection matrix C$^e$ is computed as $1 -$ CPTE$^e$. Subsequently, the C$^e$ matrix undergoes binarization through a threshold $th$ (ranging from 0 to 1 with a 0.1 step), where the connection C$^e$ (n$_i$, n$_j$) is set to 1 if CPTE$^e$ (n$_i$, n$_j$) $<=$ $th$ or 0 if CPTE$^e$ (n$_i$, n$_j$) $>$ $th$. Since CPTE$^e$ is a measure of connectivity (inversely propotional), channels n$_i$ and n$_j$ are considered “connected” if CPTE$^e$ (n$_i$, n$_i$) $<=$ $th$. The connectivity matrix averaged over epoch and case is utilized to plot the connectivity matrix of each case (AD, FTD). Given a threshold $th$, the number of active connections in the network $N_a$($th$) is estimated from the number of 1s in C$^e$. The network density $N_d$ ($th$) can now be computed as 
\begin{equation}
    N_d(th) = \frac{N_a(th)}{N_p(th)}
    \label{eq4}
\end{equation}
where $N_p(th)$ represents the total number of potential connections in the network, calculated as $\frac{n(n-1)}{2}$ = $\frac{19 (19 - 1)}{2} = 171$. Finally, $N_d$ for each case is estimated for the threshold $th$ under consideration. This procedure is replicated for the brain waves within the specified frequency bands.

The comprehensive methodological process flow for analyzing differences in large-scale brain network organization between neurodegenerative disease groups (AD, FTD) through a CPTE-based complex network model is depicted in Figure \ref{block_Diagram}.

\label{sec:Results}


\begin{figure*}
    \centering
    \includegraphics[width = 1\textwidth]{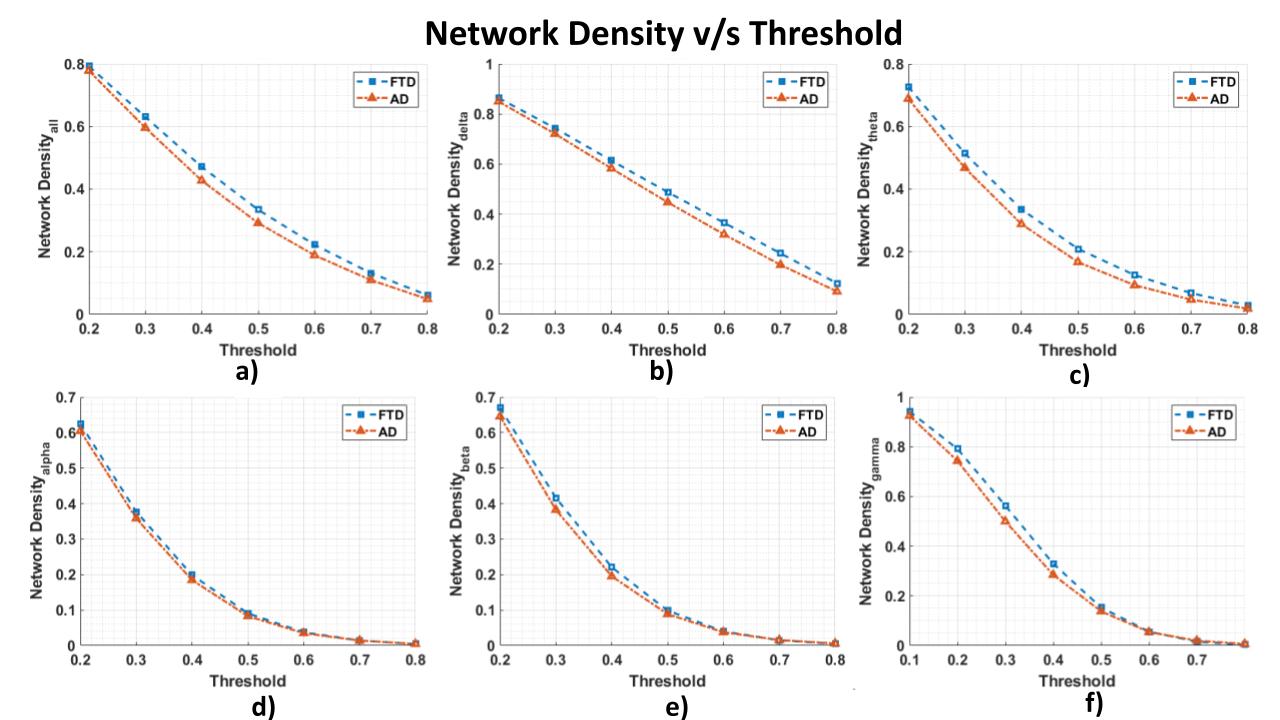}

    \caption{The figure illustrates how the connectivity density (\( N_d \)) of both the FTD and AD groups changes with varying threshold values across different frequency bands. Specifically, it highlights that the \( N_d \) value for AD is consistently lower compared to FTD across all frequency bands. Here, Figure a) all (0.5-44 Hz); b) delta (0.5-4 Hz); c) theta (4–8 Hz); d) alpha (8–13 Hz); e) beta (13–30 Hz); f) gamma (31–44 Hz) cases}
    \label{fig:ND_vs_th}
\end{figure*}

\begin{figure*}[h!]
    \centering
    \includegraphics[width=0.7\textwidth]{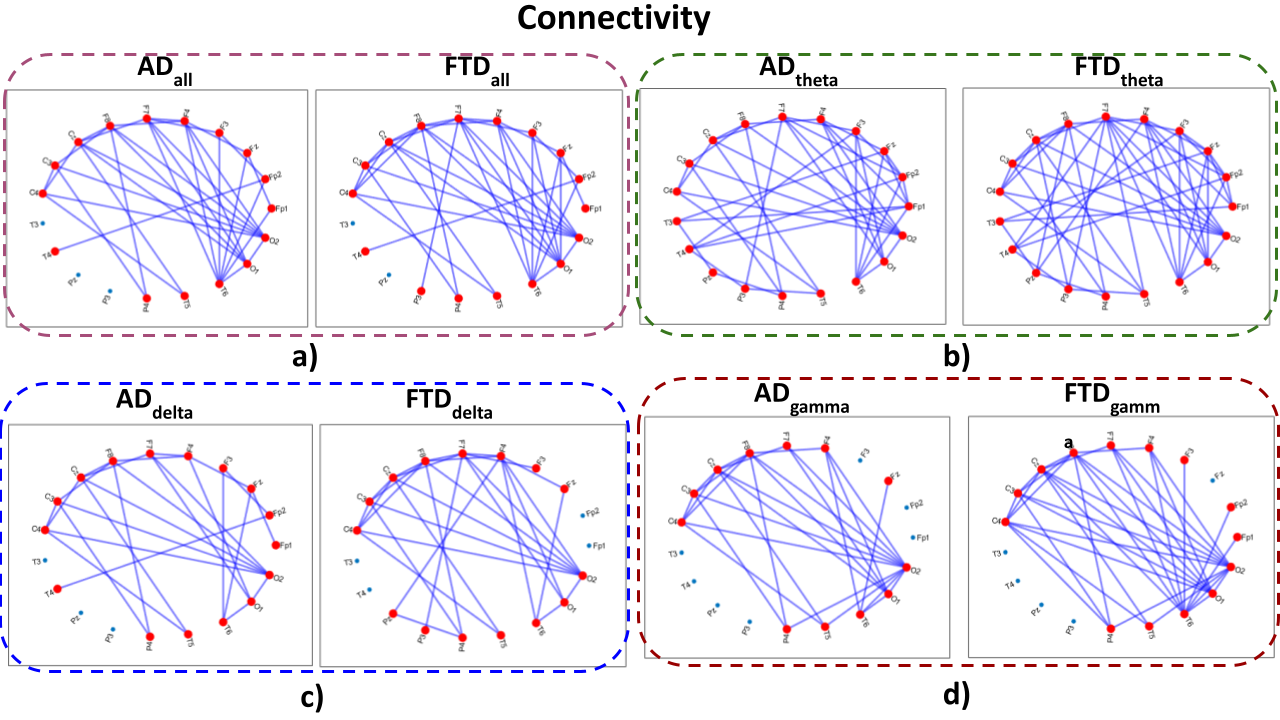}
    \caption{The figure illustrates connectivity plots between AD and FTD groups based on the synchronization measure captured by the CPTE connection matrix. It specifically focuses on cases where there was a notable difference in the \(N_d\) value between AD and FTD. Figure a) all (0.5–44 Hz); b) theta (4–8 Hz); c) delta (0.5–4 Hz); d) alpha gamma (31–44 Hz) cases}
    \label{fig:connecitivity}
\end{figure*}

 \begin{figure*}[h]
    \centering
    \includegraphics[width = 0.8\textwidth]{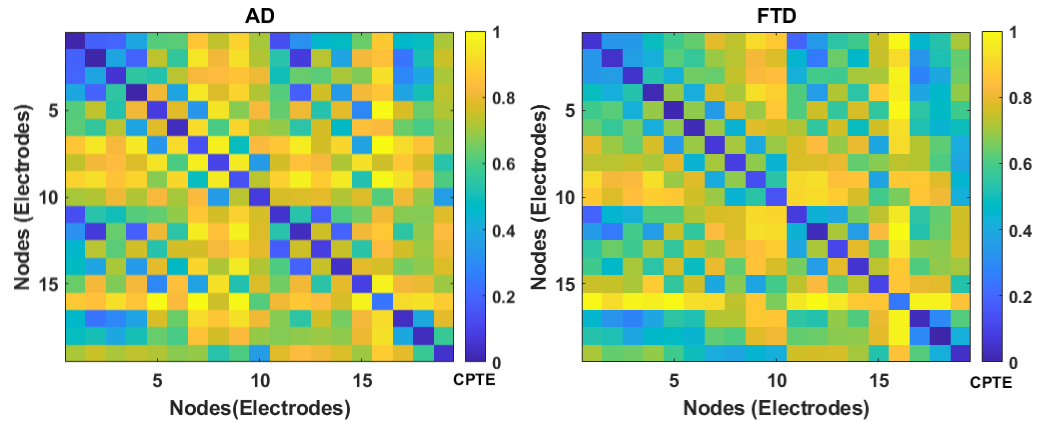}
    \caption{Illustration of the average CPTE connection matrix (all frequency band cases: 0.5 to 44 Hz) across all subjects and epochs for both the AD and FTD groups. Lower CPTE value indicates higher inter-channel synchronization.}
    \label{fig:CPTE}
\end{figure*}

     \begin{figure}[h!]
    \centering
    \includegraphics[width = 0.48\textwidth]{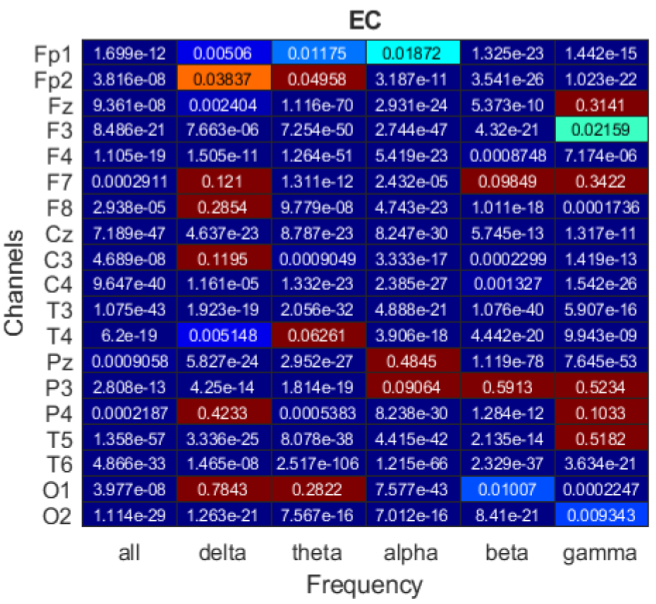}
    \caption{ The figure illustrates the p-values of the EC network parameters derived from the CPTE connection matrix of both AD and FTD groups across all channels and for all frequency band cases.}
    \label{fig:channel_bands_pvalue}
\end{figure}

\section{Results}
\label{result}
\subsection{Complex Network Analysis}
Given an EEG recording and an epoch \( e \) (for every subject, there are 150 epochs of 4s each), a synchronization matrix CPTE\(^e\) is computed, as detailed in Section \ref{CPTE_Analysis}.  The graph analysis described in Section \ref{Network_Analysis} is carried out, and the network parameters CC, SC, and EC are calculated. For each network parameter, two vectors, CC-AD and CC-FTD, are constructed, each containing the CC values calculated for all the epochs of the related group. For 36 AD subjects, CC-AD is a vector with 5400 elements (36 \(\times\) 150), whereas CC-FTD is sized 3450 (23 \(\times\) 150). Vectors SC-AD, SC-FTD, EC-AD, and EC-FTD are constructed similarly. Table \ref{feature_descrip} displays the median values of the numerical data for CC, SC, and EC in both the groups. A difference between the two groups can be observed. To validate this observation, a statistical comparison is conducted using the t-test, and the performance metrics of network parameters (CC, SC, and EC) are investigated as described in Section \ref{Network_Analysis} using RF and kNN classifiers. It may be noted from the table \ref{feature_descrip} that the medians for all the parameters are significantly different from each other in "all (0.5 - 44 Hz)" and the gamma range with \( p < 0.001 \). In particular, the AD group was found to have a higher median CC and a lower median SC and EC than the FTD group. The CC, SC, and EC parameters were able to distinguish the two groups with classification accuracy (CA) of 84.23\%, 85.85\%, and 85.02\%, respectively, as shown in Table \ref{CA} (row "all (0.5 - 44 Hz)"). These performance metrics were evaluated at the optimal threshold value of 0.6, where the CA of RF was the highest, as illustrated in Figure \ref{CA_threshold}.

For each subject and epoch \( e \), the connectivity density analysis, as detailed in Section \ref{connectivity_analysis}, is conducted by examining the synchronization matrix CPTE\(^e\) for that epoch. The analysis is presented for the entire frequency range (0.5-
44 Hz), followed by a separate frequency analysis in the ensuing section. Figure \ref{fig:ND_vs_th}a) displays the mean \( N_d \) versus \( th \) for both the AD and FTD subject groups. Overall, the \( N_d \) of the AD group was lower than that of the FTD group.  This observation aligns with the hypothesis that AD entails a process of neuronal disconnection, with FTD exhibiting a less pronounced extent of neural disconnection compared to AD \cite{neary1998frontotemporal,de2009functional}. The connectivity plot depicted in Figure \ref{fig:connecitivity} a) for both AD and FTD cases demonstrates this relationship consistently.

\subsection{Frequency Analysis}
To comprehend how brain waves in various frequency bands capture deviations between the AD and FTD groups, the proposed complex network analysis, as detailed in Section\ref{methodology}, is implemented for each band individually. This involved replacing the synchronization matrix CPTE\(^e\) of all bands (0.5–44 Hz) with specific frequency bands. In particular, the delta, theta, beta, and gamma bands of the FTD group exhibited significantly different network parameters (\( p < 0.05 \)) from AD. The CC and SC network parameters for the alpha band did not show a significant difference because of the high p-values of 0.22 and 0.59, respectively, as reported in Table \ref{feature_descrip}. Hence, alpha band is not considered for further analysis. 

The connectivity density analysis is presented in Figure \ref{fig:ND_vs_th} for all the frequency bands. It may be noted that \( N_d \) of the AD group is lower than that of the FTD group in every band but significantly differs in delta, theta, and gamma bands. The gamma band presents the highest CA of 92.87\% in distinguishing between the AD and FTD groups when all network parameters are considered, as reported in Table \ref{CA}. The performance evaluation for each band is conducted at specific threshold values, as determined from Figure \ref{CA_threshold}.


\section{Discussion}
\label{discussion}

Dementia poses an escalating challenge in today's aging society, with FTD emerging as a prominent cause of early-onset dementia alongside AD. Misdiagnosis rates for FTD are estimated at around 40$\%$, with a longer diagnostic journey compared to other dementias, particularly AD. FTD and AD exhibit distinct characteristics at baseline, progression rates, and treatment responses. Improved diagnostic tools, particularly for early detection, are crucial for enhancing treatment outcomes. EEG holds promise in this regard due to its speed, affordability, and patient tolerability, although its potential in dementia remains largely untapped. Prior studies have shown alterations in brain network organization in AD, attributed to factors such as plaque deposition and neuronal loss. Some studies have explored brain network organization in AD and FTD using complex network methods applied to EEG signals. Recently, CPTE has emerged as a computationally efficient novel parameter for characterizing synchronization between short time series, demonstrating success in investigating inter-channel interactions under various auditory stimulation conditions.

This study utilizes CPTE-based complex network analysis on EEG data to uncover insights into resting-state EEG and identify differences between AD and FTD groups. The normalized CPTE matrix derived from EEG signals is presented in Figure \ref{fig:CPTE}.  


A statistical difference between the two neurodegenerative conditions is additionally presented in Figure \ref{fig:channel_bands_pvalue} using p-values for the EC network parameter. The p-value being $<$ 0.05 in most of the cases (all of the cases when the entire frequency is used), it supports the potential of CPTE values as a parameter to differentiate between AD and FTD and analyze differences in brain network organization.

The complex network analysis of brain waves in various frequency bands using CPTE reveals significantly different network parameters (CC, SC, EC) between the two groups (p $<$ 0.05) except for alpha band as shown in Table \ref{feature_descrip}. In general, the AD group was found to have lower median CC, SC, and EC than the FTD group in the delta and theta. The findings support the idea that in AD, particularly within the delta and theta frequency bands, there is a gradual decrease in the efficiency of the brain network organization. \cite{frantzidis2014functional}, while in the remaining bands, the EC and CC network parameters showed intermediate characteristics between AD and FTD. A higher measure of connectivity is observed in all, delta, theta, and gamma frequency bands of FTD when compared with AD. It is evident from higher \(N_d\) throughout \(th\) in Figure \ref{fig:ND_vs_th} and their respective connectivity plot in Figure \ref{fig:connecitivity}. It may be observed from Table \ref{CA}, the CPTE-based network parameters significantly classify the two groups with an accuracy of 87.58\% when the entire frequency range is considered. In specific frequency bands, the gamma band demonstrates the highest accuracy of 92.87\% in discriminating between AD and FTD. The classification performance also highlights the significance of SC and EC network parameters over CC. However, EC exhibits slightly higher classification performance compared to SC. Thus, AD and FTD patients exhibit dissimilar resting-state functional brain network organizations \cite{de2009functional}.



Additionally, analyzing network characteristics across the 150 epochs of each subject reveals similar network parameters within the same subject, indicating stable network parameters during resting-state recordings. This suggests that intrasubject longitudinal differences may be more significant than intersubject cross-sectional differences. It offers insight into how pathology affects individual brain network organization over time and relates to cognitive decline.

Future objectives include recruiting a significant number of AD and mild cognitive impairment (MCI) subjects for longitudinal monitoring through dense electrode-montage EEG to investigate spatial-temporal changes in brain-electrical connectivity. Recent studies have shown promise in evaluating longitudinal changes in the EEG recordings of AD patients. These studies utilized NeuCube, a spatio-temporal data machine \cite{capecci2015feasibility, capecci2016longitudinal}. NeuCube, complemented with the proposed CPTE-based methodology, could provide deep insights into brain-electrical connectivity deterioration caused by AD, normal aging, or other neurodegenerative brain disorders.

\section{Conclusion}
\label{conclusion}

In this paper, a complex network study is carried out to compare the characteristics of the brain network organization in subjects with AD and FTD by analyzing their EEG recordings. A cohort of 36 AD and 23 FTD patients is examined using publicly available eye-closed resting-state data. The CPTE is utilized as a measure of synchronization between EEG signals. It is then used to build connectivity matrices for the complex network model. The proposed CPTE-based complex network analysis reveals significant differences in network parameters (CC, SC, EC) between AD and FTD groups, except for the alpha band (CC, SC). In particular, AD patients exhibit lower median CC, SC, and EC values compared to FTD patients in the delta and theta bands, suggesting a gradual decrease in brain network efficiency in AD. 
Additionally, the study reveals lower connectivity in AD than FTD, especially in the delta, theta, and gamma frequency bands. This indicates less efficient information exchange between brain areas, supporting the disconnection or neuronal degeneration hypothesis of AD. The CPTE-based network parameters effectively classify the two groups with an accuracy of 87.58\%, with the gamma band demonstrating the highest accuracy of 92.87\%. The classification results highlight the significance of SC and EC network parameters over CC, with EC exhibiting slightly higher classification performance compared to SC. The study demonstrates that CPTE-based graph analysis on EEG data from AD and FTD patients uncovers differences in brain network organization. This approach has the potential to identify distinct characteristics and provide insights into the underlying pathophysiological processes of various dementia forms.

\section*{Acknowledgment}
This research work was supported in part by IIT Mandi
iHub and HCI Foundation India with project number
RP04502G

 \bibliographystyle{ieeetr}  
 \bibliography{BSPC_v1_main}

 \end{document}